\documentclass[Times,4pt,aps,pra,twocolumn,showpacs,amsmath,amssymb,floatfix,footinbib,superscriptaddress]{revtex4-1}
\usepackage{amsmath,natbib,graphicx,subfigure,amssymb,graphics,amsmath,mathrsfs,CJK,color}
\usepackage{multirow,fancyhdr,color,bm,tabularx,psfrag,geometry,dcolumn,datetime}
\usepackage[colorlinks=true,linkcolor=blue,urlcolor=blue,citecolor=blue]{hyperref}
\usepackage[mathlines]{lineno}
\def\footnoterule{\kern -1mm \hrule width 5.8cm \kern 2.2mm}
\usepackage{tikz,xcolor,hyperref}
\definecolor{lime}{HTML}{A6CE39}
\DeclareRobustCommand{\orcidicon}{%
    \begin{tikzpicture}
    \draw[lime, fill=lime] (0,0)
    circle [radius=0.16]
    node[white] {{\fontfamily{qag}\selectfont \tiny ID}};\draw[white, fill=white] (-0.0625,0.095)
    circle [radius=0.007];
    \end{tikzpicture}
    \hspace{-2mm}}
\foreach \x in {A, ..., Z}
{\expandafter\xdef\csname orcid\x\endcsname{\noexpand\href{https://orcid.org/\csname orcidauthor\x\endcsname}{\noexpand\orcidicon}}}

\begin{document}

\title{2D isotropic negative permeability in a $\Lambda$-type three-level atomic system }
\thanks{ Supported by National Natural Science Foundation of China (NSFC) (Grant No.61205205) and the Foundation for Personnel training projects of Yunnan Province(grant No.KKSY201207068) of China.
}

\author{Shuang-Ying Zhang}
\email[Corresponding author: ]{zhaosc@kmust.edu.cn.}
\affiliation{Faculty of science, Kunming University of Science and Technology, Kunming, 650093, PR China}
 
\author{Shun-Cai Zhao \hspace{-1.5mm}\orcidA{}}
\email[Corresponding author: ]{zhaosc@kmust.edu.cn.}
\affiliation{Faculty of science, Kunming University of Science and Technology, Kunming, 650093, PR China}

\author{Ai-Ling Gong }
\affiliation{Faculty of science, Kunming University of Science and Technology, Kunming, 650093, PR China}
 
\begin{abstract}
A approach for two-dimensional(2D) negative
permeability in a $\Lambda$-type three-level atomic system
interacting with a probe magnetic and the superposition of two
orthogonal standing-wave fields is proposed. Through the theoretical
analysis and numerical simulation, two equally and tunable peak
maxima of negative magnetic responses are observed in the x-y plane,
and around the peak maxima region the negative permeability is
isotropic. A new avenue to 2D isotropic negative
permeability in isotropic applications via quantum optics method is achieved in our schme.
\end{abstract}

\maketitle
\section{Introduction}

The artificial metamaterials\cite{1} with simultaneously negative electric
permittivity and magnetic permeability are called left-handed materials (LHMs).
LHMs have attracted more and more attention\cite{2,3} because of their
surprising and counterintuitive electromagnetic and optical properties\cite{4,5,6}, such
as subwavelength resolution, reverse Doppler effect and Cherenkov
radiation, negative Goos-H$\ddot{\text{a}}$nchen shift, reversed
circular Bragg phenomenon, and et.c. Up to now, several approaches have been carried
out to realize LHMs in experiments, including artificial composite metamaterials
\cite{2}, photonic crystal structure\cite{7,8,9}, chiral material\cite{10,11}.
However, the left-handedness(LH) in these approaches based on spatially periodic
structure is usually anisotropic and accompanies a large absorption.

In that case, photonic resonant materials\cite{12,13,14,15,16,17,18,19,20,21,22,23} based on quantum optics
were believed to the candidate of isotropic LH without absorption. With the
help of the electromagnetically induced transparency (EIT) effect and
the Lorentz-Lorenz local field effect, the dense gases of atoms can realize LH with
low absorption, even zero absorption\cite{17,18}.
However, the magnetic permeability $\mu_{r}$ of photonic resonant materials is always close
to the free space value\cite{24} because the typical
transition magnetic-dipole moments are smaller than transition electric dipole moments
by a factor of the order of the fine structure constant ($\alpha\approx\frac{1}{137}$)\cite{12,25,26},
thus the permeability $\mu_{r}$ is difficult to get negative value.

In this literature, the aim of two-dimensional (2D) isotropic negative permeability in the dense atomic
gases under proper conditions is implemented. Although several schemes\cite{27,28,29,30,31,32,33,34,35}
for 2D LHMs have been proposed recently, the proposed 2D LHMs are based on classical
electromagnetic theory and limited by the spatially periodic structure of the medium.
Here, we firstly explore the possibility of implementing 2D negative magnetic effect based
on quantum optics with two orthogonal standing-wave fields via density matrix theory depicting
the interaction of light with matter semiclassically. And the orthogonal standing-wave field is
obtained from the superposition of the two standing-wave fields along the x and y
directions, respectively.

\section{ Theoretical model and approach}

\begin{figure}[htp]
\center
\includegraphics[scale=0.5]{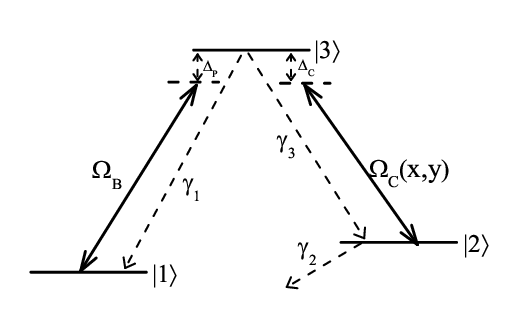}
\caption{Schematic diagram of a three-level
atomic system. The magnetic-dipole transition
$|1\rangle$$\leftrightarrow$$|3\rangle$ is excited by probe magnetic
field ($\text{\textbf{B}}$) with Rabi frequency $\Omega_{\text{B}}$. The
transition from $|2\rangle$$\leftrightarrow$$|3\rangle$ is driven by
two superposition standing-wave fields with Rabi frequency
$\Omega_{\text{C}}(x,y)$, which are along the x and y directions,
respectively.\label{1}}
\end{figure}
We consider a $\Lambda$-type three-level atom having an upper
energy level $|3\rangle$ and two closely lying lower energy levels
$|1\rangle$ and $|2\rangle$, showed in Fig.1. The magnetic-dipole
transition $|1\rangle$$\leftrightarrow$$|3\rangle$ is excited by the
magnetic component of a probe magnetic field $\textbf{B}$ with frequency
$\nu_{B}$ and Rabi frequency of $\Omega_{B}=$\textbf{B}$ \mu_{31}/2\hbar$,
where $\mu_{31}$ is the corresponding magnetic dipole moment
(magnetic-dipole matrix element). A standing-wave field
${\text{E}}(x,y)$ with frequency $\nu_{c}$ and Rabi frequency
$\Omega_{c}(x,y)$ drives the transition
$|3\rangle$$\leftrightarrow$$|2\rangle$. And the standing-wave field
${\text{E}}(x,y)$ is the superposition of two orthogonal
standing-wave fields, i.e., one is in the x direction and the second
is along y direction. However, each of the standing-wave fields is
again the superposition of two standing-wave fields along the
corresponding directions\cite{36}. Note that the parity of level
$|2\rangle$ is different from those of $|1\rangle$ and $|3\rangle$,
so the level pairs $|1\rangle$$\leftrightarrow$$|3\rangle$ and
$|2\rangle$$\leftrightarrow$$|3\rangle$ can be coupled to the probe
magnetic field and the control electric field,
respectively. $\gamma_2$ and $\gamma_3$ denote spontaneous emission
decay rates and $\gamma_1$ the collisional dephasing rate.
$\Delta_{\text{P}}$ and $\Delta_{\text{C}}$ is frequency detuning corresponding to
the probe magnetic field and standing-wave field, respectively.

In experimental investigation, the sample of metallic alkali atoms
(e.g. Na, K, Rb) might be a good candidate for our scheme. Because
the physical realizations for $\{|1\rangle$,$|2\rangle$,$|3\rangle\}$ of
the three-level system could be $\{3^{2}S_{1/2}, 3^{2}P_{1/2},
4  ^{2}S_{1/2}\}$ in neutral Na \cite{22,23}.
It should be noted that here the atoms are assumed to be nearly stationary and hence
any Doppler shift is neglected. To discuss the steady case of the
optical properties of the atomic vapor, we calculate its magnetic
permeability in two-dimensional x-y plane.

According to Schr{\"o}dinger equation, the equation of motion of the
off-diagonal density matrix elements of the atomic system are given
as follows
\begin{eqnarray}
&\dot{\rho}_{31}=&i \left(\rho_{11}-\rho_{33}\right)\Omega_{\text{B}}+i \rho_{21}\Omega_{\text{C}}(x,y)   \nonumber \\
                 &&-\rho_{31}\left(\frac{1}{2}\left(\gamma_3+\gamma_1\right)+i\Delta_{\text{P}}\right),  \label{eq1}\\
&\dot{\rho }_{21}=&-i \rho_{23}\Omega_{\text{B}}+i\rho_{31}\Omega_{\text{C}}^*(x,y)    \nonumber \\
                 &&-\rho_{21}\left(\frac{\gamma_2}{2}+i(\Delta_{\text{P}}-\Delta_{\text{C}})\right),   \label{eq2}\\
&\dot{\rho}_{32}=&i\rho_{12}\Omega_{\text{B}}+i\left(\rho_{22}-\rho_{33}\right)\Omega_{\text{C}}(x,y)  \nonumber \\
                 &&-\rho_{32}\left[\frac{1}{2}\left(\gamma_2+\gamma_3+\gamma_1\right)+i\Delta_{\text{C}}\right],  \label{eq3}
\end{eqnarray}
In the steady state case, all the atoms remain in the ground and the
atomic population in level $|1\rangle$ is close to unity. Thus we
obtain
\begin{eqnarray}
\rho_{11}=1,\rho_{22}=0,\rho_{33}=0.   \label{eq4}
\end{eqnarray}

This set of equations can be solved, for example, by writing the
matrix form,
\begin{eqnarray}
\dot{{\text{R}}}=-{\text{M}} {\text{R}}+{\text{A}}    \label{eq5}
\end{eqnarray}
with
\begin{eqnarray}
{\text{R}}=\left(
  \begin{array}{ccc}
   \rho_{31}\\
   \rho_{21}\\
  \end{array}
\right), {\text{A}}=\left(
  \begin{array}{ccc}
  i\Omega_{\text{B}}  \\
  0  \\
  \end{array}
\right),                        \nonumber \\
{\text{M}}=\left(
  \begin{array}{ccc}
  i\Delta_{\text{P}}+\left(\gamma_1+\gamma_3\right)/2&-i\Omega_{\text{C}}(x,y)\\
   -i\Omega_{\text{C}}^*(x,y)&i(\Delta_{\text{P}}-\Delta_{\text{C}})+\gamma_2/2\\
  \end{array}
\right).                                      \label{eq6}
\end{eqnarray}

The solution of matrix Eqs.5 yields the steady solution as follows:
\begin{eqnarray}
\rho_{31}=\frac{i\Omega_{\text{B}} \xi}{\Omega_{\text{C}}(x,y)\Omega_{\text{C}}^*(x,y)+\xi \left(\frac{\gamma_3 +\gamma_1}{2}+i\Delta_{\text{P}}\right)},   \nonumber \\
\rho_{21}=\frac{-\Omega_{\text{B}}\Omega_{\text{C}}^*(x,y)}{\Omega_{\text{C}}(x,y)\Omega_{\text{C}}^*(x,y)+\xi
\left(\frac{\gamma_3 +\gamma_1}{2}+i\Delta_{\text{P}}\right)}.
\end{eqnarray}  \label{eq7 }
where
$$\xi=\frac{\gamma_2}{2}+i{\left(\Delta_{\text{P}}-\Delta_{\text{C}}\right)}.$$
The atomic magnetic polarizability associated with the optical
transition $|1\rangle$$\leftrightarrow$$|3\rangle$ is
\begin{eqnarray}
\chi=\frac{2 \mu_0\mu_{31}\rho_{31}}{{\text{B}}}.
\end{eqnarray}  \label{eq8 }

According to the Clausius-Mossotti relation, the relative magnetic
permeability of the atomic media be given by
\begin{eqnarray}
\mu_r=\frac{1+\frac{2}{3}{\text{N}}\chi}{1-\frac{{\text{N}}\chi}{3}},
\end{eqnarray}  \label{eq9}
where N is the atom number density. We set some typical parameters
$\gamma_1=0.5\gamma {\text{Hz}}$, $\gamma_3=1.5\gamma {\text{Hz}}$,
$\gamma_2=\gamma {\text{Hz}}$. All the parameters are reduced to
dimensionless units by scaling with $\gamma$, where $\gamma=10^7$.
The Rabi frequency of the standing-wave field is set in the form
\begin{eqnarray}
\Omega_{\text{C}}(x,y)=\Omega_{\text{C0}}[\sin({\text{k}}x+\phi)+\sin({\text{k}}y+\varphi)],
\end{eqnarray}  \label{eq10}
where ${\text{k}}=\frac{2\pi}{\lambda}$ is wave vector and
wavelength is $\lambda=4\mu{{\text{m}}}$. Other parameters are taken
with $\mu_{31}=6.6\times10^{23}{\text{C}}
{\text{m}}^2{\text{Hz}}$, $\Omega_{\text{C0}}=5\gamma$ and atomic
concentration $\text{N}=5.0\times10^{24}{\text{m}}^{-3}$\cite{22}.

\section{Results and discussions}

In the preceding sections, we calculate the relative magnetic
permeability in present atomic vapor system. In what follows, let us
discuss its optical properties by simulating the magnetic permeability.
The behaviors of the magnetic permeability corresponding to the frequency
detunings and the initial phases will be of interest.

\begin{figure}[htp]
\center
\includegraphics[totalheight=1.0 in]{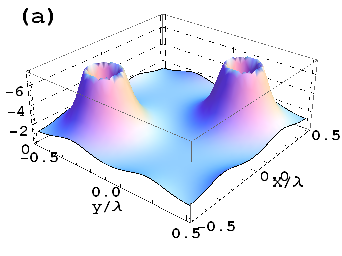}\includegraphics[totalheight=1.0 in]{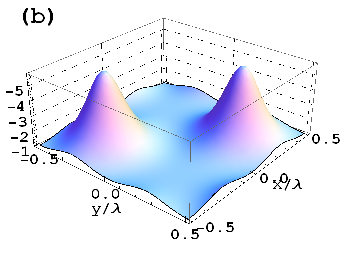}
\hspace{0in}%
\includegraphics[totalheight=1.0 in]{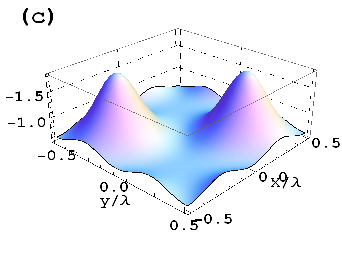}\includegraphics[totalheight=1.0 in]{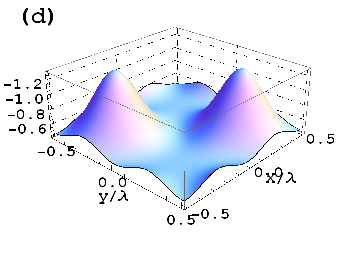}
\caption{(Color online) Plots for the 2D real part of magnetic
permeability Re[$\mu_{r}$]:  Re[$\mu_{r}$] versus x and y for different
values of probe magnetic detunings $\Delta_{p}$
(a)$\Delta_{p}$=-2.5$\gamma$,(b)$\Delta_{p}$=-4.1$\gamma$,
(c)$\Delta_{p}$=-7.5$\gamma$, (d)$\Delta_{p}$=-9$\gamma$. And the
initial phases $\phi$=0, $\varphi$=0 and $\Delta_{c}$=6.3$\gamma$ were taken.\label{2}}
\end{figure}

The calculated real parts of magnetic permeability Re[$\mu_r$] for
different probe detuning parameters $\Delta_{p}$ are shown in Fig.2 in a
three-dimensional Cartesian coordinate system.
The behavior of negative magnetic response is strong when the probe detunings are set
$-2.5\gamma$, $-4.1\gamma$, $-7.5\gamma$ and $-9\gamma$ in Fig.2(a), (b), (c) and (d), respectively.
As shown in Fig.2(a), two circular crater-like patterns with the peak
maxima being about -6 locate in the quadrants I and III.
However, the peak patterns of Re[$\mu_r$] change into two circular spike-like patterns
with their values being about -5 in Fig.2(b). And the peak values of the two circular spike-like patterns
collapse to -2 when the value of $\Delta_{p}$ is tuned to 7.5$\gamma$ in Fig.2(c).
The peak values of the two circular spike-like patterns continue to decrease to -1.2 when $\Delta_{p}$ is
set to -9$\gamma$ in Fig.2(d). We noted the circular distribution of Re[$\mu_r$] has the same value circling around the center
of the crater-like or spike-like patterns in Fig.2(a), (b), (c) and (d). The circular equivalent distribution shows the isotropic Re[$\mu_r$] in the x-y plane.

\begin{figure}[htp]
\center
\includegraphics[totalheight=1.0 in]{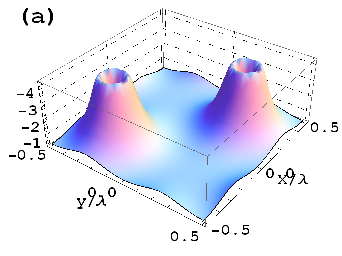}\includegraphics[totalheight=1.0 in]{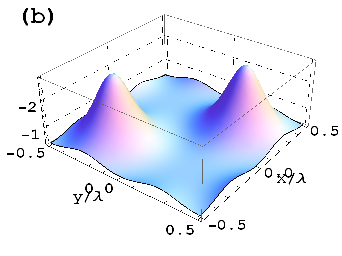}
\hspace{0in}%
\includegraphics[totalheight=1.0 in]{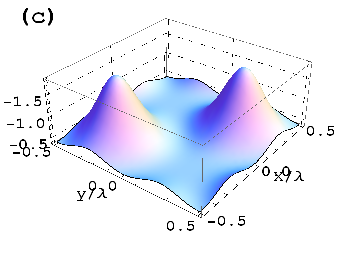}\includegraphics[totalheight=1.0 in]{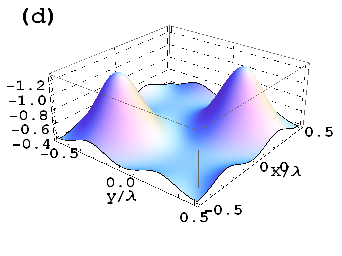}
\caption{(Color online) Plots for the 2D real part of magnetic
permeability Re[$\mu_{r}$]: Re[$\mu_{r}$]  versus x and y for different
 values of the standing-wave field detuning $\Delta_{c}$. (a)$\Delta_{c}$=-4.5$\gamma$,
 (b)$\Delta_{c}$=-1.9$\gamma$, (c) $\Delta_{c}$=0.2$\gamma$,
 (d)$\Delta_{c}$=3.5$\gamma$. $\Delta_{p}$=-10$\gamma$, and all other parameters are the same as those in Fig.2.\label{3}}
\end{figure}

Another tunable parameter is the coupling standing-wave fields detuning $\Delta_{c}$ from the transition
$|2\rangle$$\leftrightarrow$$|3\rangle$, and the corresponding 2D plots are shown in Fig.3.
In Fig.3, the values of the standing-wave fields detunings $\Delta_{c}$
equal to (a) -4.5$\gamma$, (b) -1.9$\gamma$, (c) 0.2$\gamma$, and (d)
3.5$\gamma$. Comparing with Fig.2, the behavior of negative magnetic response is relatively weak via the
tuning detuning $\Delta_{c}$. The peak maxima in the center of quadrants I and III changes
with not only the pattern but also the values. Two circular crater-like patterns with their peak values being -4 when $\Delta_{c}$ is tuned
to -4.5$\gamma$ in Fig3.(a). The circular crater-like patterns transform rapidly into the circular
spike-like patterns in Fig.3 from (b) to (d), and their peak values show a downward trend, i.e., from -2.6, -1.8, to -1.2.
It's also noted that the peaks of Re[$\mu_r$] are circular distribution around their centers from the crater-like
or spike-like patterns. The circular distribution of Re[$\mu_r$] shows the isotropy of Re[$\mu_r$] in the x-y plane.

\begin{figure}[htp]
\center
\includegraphics[totalheight=1.0 in]{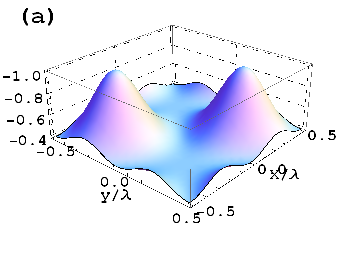}\includegraphics[totalheight=1.0 in]{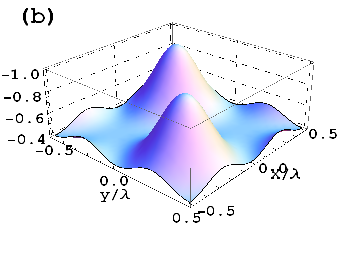}
\hspace{0in}%
\includegraphics[totalheight=1.0 in]{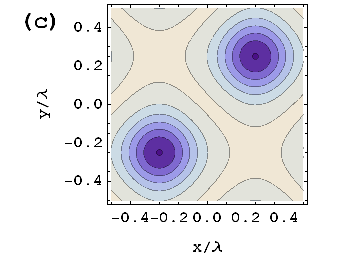}\includegraphics[totalheight=1.0 in]{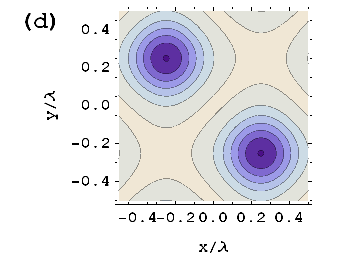}
\caption{(Color online) Plots for the 2D real part of magnetic
permeability Re[$\mu_{r}$]: Re[$\mu_{r}$] versus x and y for different
initial phases. (a) $\phi$=0, (b) $\phi$=-$\pi$, and their corresponding contour plots shown in (c), (d).
$\Delta_{p}$=-10$\gamma$, $\Delta_{c}$=7.2$\gamma$, all other parameters
are the same as those in Fig.2.\label{4}}
\end{figure}

In the follows, the behavior of Re[$\mu_r$] dependent different
initial phases is discussed in Fig.4(a) and (b). Fig.4 (c) and (d)
are the contour plots corresponding to Fig.4(a) and (b), respectively.
In Fig.4 (a), the initial phases for the standing-wave field
are set $\phi$=0, $\varphi$=0 against $\phi$=-$\pi$, $\varphi$=0 in Fig.4 (b).
We noticed that the amplitude ranges of Re[$\mu_r$] both are [-0.4, -1.0] and the
two spike-like distribution graphs remain the same in Fig.4(a) and (b).
The difference tuned by $\phi$ between Fig.4(a) and (b) is the position.
The positions of the two spike-like distribution graphs are in quadrants I and III
in Fig.4(a) when $\phi$=0, while the positions change to quadrants II and
IV when $\phi$=-$\pi$ in Fig.4(b). The variation in positions is also displayed by their
corresponding contour plots in Fig.4 (c) and (d), respectively. The contour plots corresponding to
the two spike-like distribution graphs are concentric circles shown by Fig.4 (c) and (d).
Taking the circle center as center, all of the values of Re[$\mu_r$] on the concentric circles are the same.
The same values of Re[$\mu_r$] in the directions of $360^{0}$ demonstrate the isotropy.
So, an isotropic and homogeneous negatively permeability is obtained in this atomic system. And the
$\Lambda$-type three-level atomic system maybe particularly essential for designing
devices such as a sub-wavelength focusing
system or negative-index super-lens for perfect imaging.

\section{Conclusion}

In conclusion, we presented a theoretical scheme for 2D isotropic negative
permeability in the dense atomic gas via two orthogonal standing-wave fields.
By adjusting the detunings corresponding to the probe and two orthogonal
standing-wave fields, the flexible control negative magnetic
responses are observed in the x-y plane, and around the center of the
circular crater-like or spike-like patterns Re[$\mu_r$] has the same value
, which demonstrates the isotropic Re[$\mu_r$]. When the initial phases
in the standing-field are manipulated, the positions of peak changes from
quadrants I and III to quadrants II and IV, and their contour plots demonstrate
clearly the isotropic and homogeneous Re[$\mu_r$].
The atomic vapor system in our scheme could
be the candidate for 2D negative permeability and be a new avenue to
2D negative refraction research. Compared to other
proposal, the atomic vapor scheme can eliminate some manufacturing
constraints as well as reducing experimental difficulties. And we hope our
scheme for 2D isotropic negative permeability could achieved in the coming
atomic medium experimentally.

\end{document}